\documentclass[twocolumn,aps]{revtex4}

\usepackage{graphicx}% Include figure files
\usepackage{dcolumn}% Align table columns on decimal point
\usepackage{bm}% bold math

\begin{document}
%\preprint{APS/123-QED}

\title{Complete stabilization and improvement of the characteristics of tunnel junctions by thermal annealing}
% Force line breaks with \\

\author{P. J. Koppinen}
%\email{panu.koppinen@phys.jyu.fi}
\author{L.M. V\"aist\"o}%
\author{I.J. Maasilta}
\affiliation{%
Nanoscience Center, Department of Physics, University of Jyv\"askyl\"a, P. O. Box 35, 
FIN--40014 University of Jyvä\"askyl\"a, Finland.
}

\date{\today}% It is always \today, today,

\begin{abstract}
We have observed that submicron sized Al--AlO{$_x$}--Al tunnel junctions can be stabilized completely by annealing them in vacuum at temperatures %%@
between  $350^{\circ}$C and
$450^{\circ}$C. In addition, low temperature characterization of the samples after the annealing treatment showed a marked improvement of the %%@
tunneling characteristics due to disappearance of unwanted resonances in the current.  Charging energy, tunneling resistance, barrier
thickness and height all increase after the treatment. The superconducting gap is not affected, but supercurrent is reduced in accordance with the %%@
increase of tunneling resistance.  
\end{abstract}

%\pacs{}% PACS, the Physics and Astronomy
\maketitle
Submicron sized metal--insulator--metal (MIM) tunnel junctions have been used widely in  applications such as radiation detectors\cite{enss}, 
single electron transistors (SET)\cite{NATO}, and tunnel junction coolers and thermometers \cite{SINIS}. They are also the central element in %%@
superconducting quantum bits \cite{sq1,sq2,sq3,sq4,sq5}, and magnetic junctions \cite{mood} are developed for memory applications. However,  there are %%@
often
problems with stability of the junctions in room air conditions, the most noticable being aging, i.e. a slowly creeping increase of the tunneling %%@
resistance $R_T$ with time.
Instability of the junctions is an obvious drawback when considering practical and commercial applications of tunnel junctions. 

One possible reason for aging is the absorption of unwanted molecules into the barrier. Direct knowledge of these processes is sketchy, although it is %%@
known that vacuum conditions can slow down aging significantly. \cite{Nahum}  Another possibility is the relaxation of the oxide barrier properties. %%@
Typically, tunnel barriers in MIM junctions are fabricated by room temperature thermal oxidation of a disordered base electrode. This procedure %%@
results in formation of a polycrystalline 
or amorphous oxide layer, in which atoms are not in their global equilibrium positions. The relaxation from this "glassy" state could take months at %%@
room temperature. Moreover, in AlO$_x$ barriers studied here, chemisorbed oxygen ions bound on the barrier surface during fabrication have also been %%@
shown to be a source of instabilities. \cite{tan}  

A typical remedy for finding a better energy minimum is thermal annealing, the artificial acceleration of the relaxation processes by  heating up the %%@
sample for some time and then letting it cool again. 
For technologically important Nb/Al-AlO$_x$/Nb junctions, annealing treatments in gas atmospheres (He, N$_2$, Ar, air) were shown to change the %%@
barrier properties, always increasing $R_T$. \cite{gur,leh,oliva,cimpoiasu} On the other hand, for magnetic tunnel junctions, improvements in their %%@
performance were obtained with vacuum anneals. \cite{MTJ,wang}  

In this letter we discuss how Al--AlO$_{x}$--Al tunnel junctions can be completely stabilized, and their characteristics improved by thermal annealing %%@
in vacuum. The general trend of increase of $R_T$ with the annealing is reproduced. The bias voltage dependence of the tunneling conductance reveal %%@
that the junction characteristics become more ideal after the anneal. In addition, the charging energy, the effective barrier thickness and height all %%@
increase due to the annealing. Previous work on annealed Al-AlO$_x$-Al junctions \cite{zorin} is partly inconsistent with our results, which is not %%@
surprising, as they used a forming gas (H$_2$/N$_2$) atmosphere in their process instead of the inert vacuum used here.         

A total number of 178 Al--AlO$_{x}$--Al samples were annealed, containing either a single junction  or two junctions in series, fabricated by %%@
conventional electron beam (e-beam) and standard double angle vacuum evaporation techniques. All the junctions had either a junction area $\sim 0.05 %%@
\mu$m$^2$ or  $\sim 0.15 \mu$m$^2$.
All samples were fabricated on oxidized silicon substrates, 
which were usually cleaned with an oxygen plasma just before Al deposition. For a few samples which were not annealed, plasma cleaning was not %%@
performed to see how the aging process is affected. The 300 nm wide and 55 nm thick Al wires were e-beam evaporated in high vacuum at a rate $0.3$ %%@
nm/s, and the AlO$_{x}$ barriers were thermally grown {\it in situ} in pure oxygen at $\sim0.2$-$0.5$ mbar for $3-5$ minutes. To reduce the effects of %%@
contamination, 
all samples were post-oxidized {\it in situ} as a last step in $\sim 0.5$ bar of pure oxygen for five minutes.
The samples were annealed in a vacuum chamber ($P < 10^{-3}$ mbar) to temperatures ranging 
from $200^{\circ}$C
to $500^{\circ}$C. The temperature cycle consisted of a quick (minutes) heating phase followed by a slow (hours) cooling cycle; details can be found %%@
in Ref. \cite{AIP}.

\begin{figure}
\includegraphics[width=8.5cm]{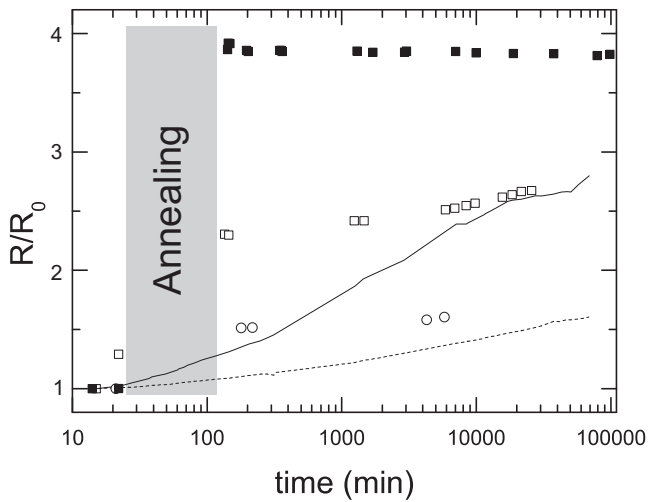}
\caption{\label{fig:aging} Aging of a non--annealed sample with (dashed) and without (solid) plasma cleaning of the substrate.  
Open circles represent aging when the sample was annealed at $200^{\circ}$C, open squares at $300^{\circ}$C and filled squares at $400^{\circ}$C, %%@
respectively. All resistances are normalized with initial resistance at time equals zero ($R_{0}$).  
}
\end{figure}

Aging was monitored before and after the annealing by measuring the sample resistance at room temperature vs. time, with all samples stored in room %%@
air.
Representative results are shown in Fig.\ref{fig:aging}. The solid and dashed lines show the typical aging of non--annealed samples with and without %%@
plasma cleaning of the substrate, respectively. The two samples shown were fabricated in the same evaporation and oxidation step to ensure identical %%@
conditions. 
It is clear that plasma cleaning of the substrate before Al deposition results in significant reduction of aging, but does not stabilize the %%@
junctions. We conclude that some organic contaminants, most likely resist residue on the substrate, are partly responsible for the aging phenomenon.
The aging for a few representative annealed samples is also shown in Fig. \ref{fig:aging}, with different symbols denoting different annealing %%@
temperatures. The annealing process started 
approximately 20 minutes after the end of fabrication, which defines the zero of time and the initial resistance $R_0=R(t=0)$ for each sample. %%@
Clearly, annealing temperatures $200^{\circ}$C and $300^{\circ}$C do not lead to stabilization, whereas complete stabilization occurs at  %%@
$400^{\circ}$C. Note that the last datapoints correspond to times over two months after the fabrication. Complete stabilization was also observed at %%@
annealing temperatures $350^{\circ}$C and $450^{\circ}$C (data not shown), but $500^{\circ}$C always resulted in breakdown of the junctions. 
In addition, the thermal treatments always led to increase of the sample resistance. At
$400^{\circ}$C, where we have the largest amount of data (120 samples),  the increase varied from a factor of two to a factor of eight, typically %%@
being between a factor of two to four. The range of initial $R_T$/junction for the $400^{\circ}$C  treatments was from 1 k$\Omega$ to 100 k$\Omega$, %%@
and 
no clear dependence of the $R_T$ increase on the initial $R_T$ was observed. Curiously, the low temperature anneals at $200^{\circ}$C and %%@
$300^{\circ}$C that did not lead to stabilization show approximately the same logarithmic aging rate as the non-annealed samples, see Fig. %%@
\ref{fig:aging}, but with a jump of the resistance caused by the treatment. This means that some of the resistance increase is not correlated with the %%@
stabilization in any way.      

\begin{figure}
\includegraphics[width=8.5cm]{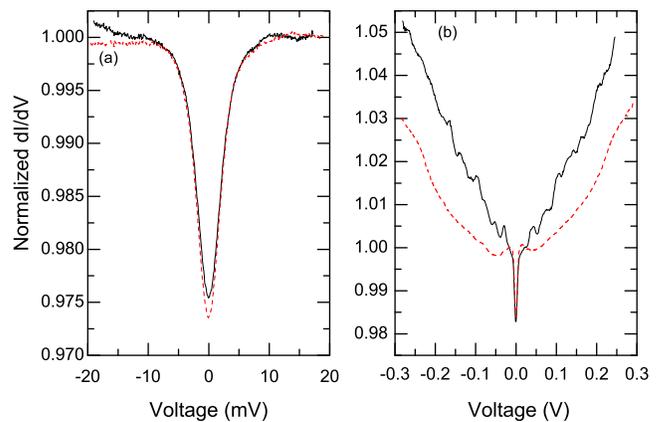}
\caption{\label{fig:GvsV} (Color online) A typical normalized $dI/dV$ vs. $V$ spectrum in units of $G_T$ of a sample before (solid) and after (dashed, %%@
red) annealing. (a) shows 
a narrow bias range focusing on Coulomb blockade, while (b) displays a wide bias voltage range (tunneling spectrum).}
\end{figure}

%\begin{figure}
%\includegraphics[width=8.5cm]{ALMFP_both3}
%\caption{\label{fig:SEM} An SEM micrograph of an Al film (a) before and (b) after annealing at $400^{\circ}$C.}
%\end{figure}

For ten two junction samples, the normal state tunneling properties were characterized before and after the annealing at $400^{\circ}$C by measuring %%@
the tunneling current and differential conductance $dI/dV$ vs. bias voltage at 4.2 K with standard lock--in techniques. 
In this case, the low bias voltage  regime ($|V| < 20$mV)  provides information on the charging energy, i.e. capacitance, of the junctions. 
The charging energy $E_{C}=e^2/(2C_{\Sigma})$, where $C_{\Sigma}$ is total capacitance of the junctions and the island, can be obtained from the %%@
height of the zero-bias Coulomb blockade (CB) dip of $dI/dV$ [Fig. \ref{fig:GvsV} (a)] when $k_BT >> E_C$, which is the appropriate limit for our %%@
junctions at 4.2 K.   
For two junctions in series, the height of the dip $\Delta G$ and $E_C$ are related by 
$\Delta G/G_{T}=E_{C}/(6k_{B}T)$
\cite{SINIS}, where $G_T$ is the tunneling conductance around $V=0$ without the CB dip.   
Figure \ref{fig:GvsV} (a) shows a typical CB dip in our samples before and after the annealing. 
Usually $E_C$ increased by $5$ -- $10$ $\%$ after the annealing, with 
initial $E_C$s ranging from 9.8 $\mu$eV to 53$\mu$eV corresponding to capacitances $C_{\Sigma}=8.2$--$1.5$ fF. 
The values of $R_T=1/G_{T}$ increased $1.5$--$3$ times for these samples after the annealing, varying initially from sample to sample between 9.5 %%@
k$\Omega$ and 151 k$\Omega$. 

In Fig. \ref{fig:GvsV} (b) we plot a typical tunneling conductance spectrum ($dI/dV$) in a wide voltage bias range $|V| < 300$ mV. The spectrum shows %%@
many resonances (peaks) before
the annealing, indicating that initially our barriers are far from perfect. These resonances are usually caused by unwanted impurities within the %%@
barrier, allowing inelastic or resonant tunneling at some specific energies \cite{wolf}. Almost all resonances have clearly disappeared after the %%@
annealing treatment, in addition the strength of the voltage dependence has been reduced. One possibility discussed before \cite{gur} is that aluminum %%@
hydrates present in the disordered barriers dehydrate in the annealing, causing the resonances due to O-H groups to disappear.  
In an idealized case, i.e. by assuming a trapezoidal barrier and that direct tunneling dominates, the voltage dependence of the tunneling conductance %%@
is commonly parametrized in the WKB approximation as 
$G=G_0[1+(V/V_{0})^2]$, where $V_{0}^2=4\hbar^2\phi_{0}/(e^2md^2)$  and $G_0=e^2A\sqrt{2m\phi_{0}}/(h^2d)\exp(-2d\sqrt{2m\phi_{0}}/\hbar)$. Here
$A$ is the junction area, $\phi_0$ is the average barrier height and $d=(m^*/m)^{1/2}d_{ox}$ is the effective barrier thickness, where $d_{ox}$ is the %%@
physical thickness and $m^*$ the barrier effective mass. \cite{wolf} 
Within this model we conclude that the curvature parameter $V_{0}$ increases in the annealing treatment. Together with the increase of
$E_{C}$ and $R_T$ we can  deduce that both the barrier height $\phi_0$ as well as its effective thickness $d$ increase with the annealing treatment. %%@
We stress that the increase of $E_C$ alone (decrease of $C_\Sigma$) does not guarantee that $d$ increases, as the dielectric constant most definitely %%@
also changes during the annealing.  
Similar kind of changes in junction parameters have also been reported earlier at least for gas annealed Nb/Al--AlO{$_x$}--Nb junctions %%@
\cite{cimpoiasu,leh}. The change in the absolute values of the barrier parameters are difficult to estimate, as the numerous resonances in the %%@
conductance spectrum before the annealing makes fitting to the model unreliable. Quadratic fit to the annealed data yields the values $d=$9.5 \AA  and %%@
$\phi_0=$1.6 eV for the sample in Fig. \ref{fig:GvsV}.  

In addition to the improved tunneling characteristics, we also observed an 
improvement in the Al film quality by the treatments. This was checked by measuring the resistance of a $70$ nm thick, 550 nm wide wire sample at %%@
$4.2$K before and after annealing at $400^{\circ}$C. The resistivity decreased by $\sim 30 \%$ from 5.1 to 3.4 $\mu\Omega$cm , corresponding to an %%@
increase in mean free path from 7.7 nm to 12 nm. High resolution SEM images of the film were also taken before and after the annealing, without any %%@
noticeable difference in the film grain structure. 

\begin{figure}
\includegraphics[width=8.5cm]{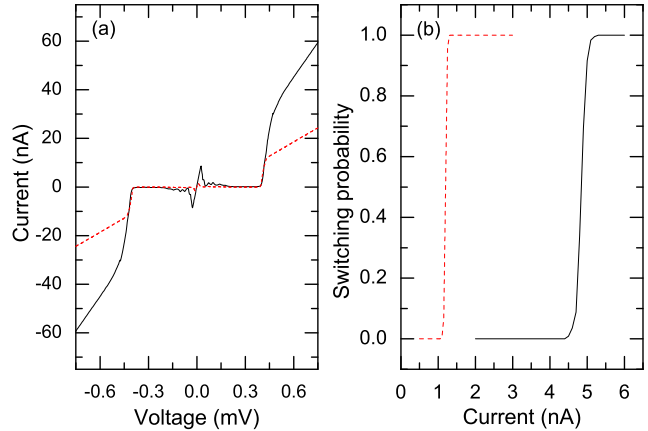}
\caption{\label{fig:jjdata}(Color online) (a) Typical IV curve and (b) switching probability characteristics of a superconducting single junction %%@
sample before (solid) and after (dashed, red) annealing measured at 80 mK.}
\end{figure}

In addition to  measurements at 4.2 K, we also determined the  
 superconducting properties of six unshunted single junction samples, cooled to below 100 mK
in a dilution refrigerator.  
A representative I-V characteristic before and after annealing at $400^{\circ}$C is presented in Fig \ref{fig:jjdata} (a). The increase of $R_T$ is %%@
apparent as the decrease of the slope at high bias, in this sample approximately a factor of three. The  superconducting gap $\Delta$ was determined %%@
from the curves, with the maxima of the $dI/dV$ giving $2\Delta$, whose value did not change after the annealing. Reduction of the critical current %%@
$I_C$ around zero bias is also visible, as expected from  the Ambegaogar--Baratoff relation, \cite{NATO} where $I_C R_T=$ const for a constant %%@
$\Delta=$. 
To study the supercurrent branch more accurately, we measured the switching currents 
by the cumulative histogram method, where a set of current pulses of constant height $I_p$ were fed through the sample at a frequency 100 Hz and a %%@
duty cycle 5 \%, and the time dependent voltage response was measured. By counting how often the junction switched to the finite voltage state at  a %%@
particular $I_p$, the switching 
probability distribution of the sample is directly determined. Fig. \ref{fig:jjdata} (b) shows the obtained results for the sample whose IV data is %%@
shown in Fig. \ref{fig:jjdata} (a). It is clear that the switching current $I_S$ (defined as the value where the probability $=0.5$)  is reduced to %%@
approximately one fifth of the initial value, while the width of the distribution also decreases.

By calculating the fluctuation free critical current $I_{C}$ from the Ambegaogar--Baratoff relation  
and estimating the capacitance of the junction as half of the capacitance measured for the two junction samples of the same junction size, the ratio %%@
of the Josephson coupling energy $E_J$ and the charging energy $E_C$ can be estimated to be $E_{J}/E_{C}\sim 1.8$. This means that the 
samples are in the underdamped regime, where for unshunted junctions the prediction is that
$I_{S}\propto R_{T}^{-3/2}$. \cite{Joyez}  The size of reduction in the switching current $I_{S}$ is therefore fully consistent with the increase in %%@
$R_{T}$.  

In conclusion, Al-AlO$_{x}$-Al tunnel junctions can be completely stabilized against aging by vacuum annealing at $350-450^{\circ}$C . This is an %%@
obvious advantage, when considering practical and commercial applications of tunnel junction devices. The improvement of
the tunneling characteristics was also apparent as  unwanted resonances disappeared after the treatment. Annealing led to a consistent reduction of %%@
junction capacitance and tunneling conductance for all samples studied, caused by an increase in the barrier height and thickness. Engineering
these sample parameters in a controlled way by thermal vacuum annealing was not yet possible, partly because the microscopic understanding of the %%@
processes taking place is unclear.  One possible application for the annealing is to increase the sensitivity of Josephson junction threshold current %%@
detectors, currently used for example in superconducting quantum bit readouts \cite{sq3} or in shot-noise measurements. \cite{jukka} With annealing, a %%@
reduction of the critical current can be achieved without a significant reduction of the capacitance, which decreases the sensitivity. \cite{jani}

We thank J. M\"annik for discussions. This work has been supported by the Academy of Finland projects No. 105258 and 205476 (TULE program). P.J.K. %%@
acknowledges National Graduate School in Materials Physics 
and Emil Aaltonen Foundation for partial financial support.

%\bibliography{annealing}

\end{document}